# Lattice-Based Ring Signature Scheme under the Random Oracle Model

Shangping Wang, Ru zhao

*(Shaanxi Key Laboratory for Network Computing and Security Technology, Xi'an University of Technology, Xi'an, 710054, China)*

**Abstract:** On the basis of the signatures scheme without trapdoors from lattice, which is proposed by Vadim Lyubashevsky in 2012, we present a new ring signature scheme from lattice. The proposed ring signature scheme is an extension of the signatures scheme without trapdoors. We proved that our scheme is strongly unforgeable against adaptive chosen message in the random oracle model, and proved that the security of our scheme can be reduced to the hardness of the small integer solution (SIS) problem by rejection samplings. Compared with the existing lattice-based ring signature schemes, our new scheme is more efficient and with shorter signature length.

*Keywords:* lattice; ring signature; random oracle model; unforgeable;

Mathematics Subject Classification (2010) 94A60; 81T25

## 1. Introduction

Recent studies show that the cryptography schemes based on the problems of large integer factoring and discrete logarithm have been unable to resist the quantum attacks, so we must research more efficient and more secure new cryptosystems to resist the quantum attacks. In the 18th century, mathematicians Lagrange, Gauss and Minkowski began to study lattice. Because lattice is of linear structure and the operation based on the lattice is mostly linear, the hard problem on lattice can be used to build new and efficient public key cryptosystems. The lattice-based cryptosystem is assumed can resist quantum attacks, so far the quantum algorithm[1] which can solve the hard problems on the lattice has not emerged in polynomial time , therefore the lattice-based cryptography has became one of the best choice in the era of quantum computing time.
Constructing new public key cryptosystems based on the hard problem in lattice has become a new research hotspot of cryptography.

### 1.1 Ring Signature

A digital signature is a kind of encryption technology, which has message integrity, authentication and non-repudiation. There has been a class of digital signatures which is known as the group digital signature. In the cryptosystem, the entity is a group, and the membership is a complex structure. A group-oriented digital signature contains various different types of digital signatures, such as group signature, ring signature, signature threshold.

    2001, Rivest et al. proposed a new signature technique called ring signature in the context of anonymously leaked secret [2]. Ring signature can be regarded as a special group signature [3]. There is no established process group, and for those verifiers, the signer is completely anonymous. Meanwhile, ring signature provides a clever method for anonymous leaking secrets. This unconditional anonymity in long-term protection of information needs some special environment. Because of its unique nature, the ring signature technique can be widely used in anonymous electronic voting, e-government and e-currencies.

    2004, Yevgeniy Dodis et al. discussed ad-hoc groups anonymous authentication issues [4], and presented a ring signature scheme based on one-way

accumulators [5]. Gan Zhi et al. used the valve channel method of embedding authentication information to construct a more efficient verifiable ring signature scheme [6]. Amit KAwasthi put forward an efficient identity-based ring signature and proxy ring signature scheme [7]. Javier Herranz et al. put forward a new identity-based ring signature scheme [8]. JiQiang Lv et al. have proposed a ring authenticated encryption scheme which is the combination of encryption and authentication signature[9] and has the verifiability. Tianjie Cao improved the group signature scheme [11]. Tony K.Chan put forward the concept of blind ring signature [12].

2005, KCLee put forward a revocable anonymous ring signature scheme [13]. Patrick P.Tsang designed an association ring signature scheme [14], which can be used for electronic voting and electronic cash systems. Lan Nguyen constructed an accumulator based on bilinear pairing identity-based ring signature scheme [15], and the signature length is fixed. Qianhong Wu et al. constructed an efficient blind ring signature scheme [16]. Qiang Lei et al. constructed an anonymous fingerprinting protocol with the ring signature scheme [17]. Joseph K.Liu et al. who focused on the key leak issues the first time described a forward security ring signature and key encapsulation ring signature scheme [18], and then extended it to the threshold scheme.

In 2006, Yiqun Chen put forward a suitable P2P network identity-based anonymous designated ring signature scheme [19]. Adam Bender analyzed the security definition of the ring signature scheme and fully taken into account the actual ability of an attacker. Eventually, he gave more reliable security definitions and proposed security in the standard model of the two ring signature schemes [20].

In 2008, Gentry, Peikert and Vaikuntanathan raised a hash-and-sign signature scheme based on the hardness of worst-case lattice problem [21]. Lyubashevsky and Micciancio built a signature scheme based on the hardness of worst-case ideal lattice problem [22]. However, the hash-and-sign signature scheme is more inefficient and with larger key length, so it has the higher storage costs.

In 2009, due to the inefficiency of the hash-and-sign signature, Lyubashevshy proposed a new digital signature scheme, which using Fiat-Shamir framework based on the hardness of worst-case ideal lattice problem [23]. The scheme was the only one using the Fiat-Shamir technique at that time. It has made a major contribution to building lattice-based signature scheme using the Fiat-Shamir technique. The first contribution of this work is adapting the ring-SIS based scheme to one based on the hardness of the regular SIS problem. The second contribution is reducing the signature length to $\tilde{O}(n)$. The third contribution is showing that the parameters of the scheme can be set so that the resulting produces shorter signatures.

2010, Wang and others, who based Cash signature scheme, proposed a lattice-based ring signature scheme against the fixed ring attack in the standard model [24]. A disadvantage of this scheme is that the dimension of the new bonsai tree should be expanded larger than the bonsai tree in the original signature, because the ring members need to join the others' public keys. That makes the ring signature scheme has larger key size and higher storage cost. But this scheme only uses a small integer with modular multiplication and modular addition, it has high computational efficiency.

In 2012, MiaomiaoTian, who added messaging technology based Boyen's scheme, proposed a ring signature scheme based on lattice [25] and gave formal security proof in the standard model. This scheme is security under adaptive chosen message attack. Although the scheme has high efficiency, its public size, private size and signature length are large, and it has high storage cost.

In 2012, Vadim Lyubashevsky presented a signature scheme without trapdoor [26] based on lattice. The scheme is provably secure in the random oracle model. Its private key, public key and signature size are much smaller than the hash-and-sign signature scheme. The scheme greatly reduces the storage cost and has great efficiency improvement.

### 1.2 Our contribution

We present a new lattice-based ring signature scheme based on Vadim Lyubashevsky's lattice signatures without trapdoors. That is, our scheme is extension of the Lyubashevsky's lattice signatures [26], we extend it to be a ring signature scheme. The extension need elaborated, in the key generation algorithm, produce the key pair for the each user by trapdoor function. In the signature algorithm, the signer uses the others' public keys and his secret key to generate the ring in the name of all members in the ring. In terms of security, the scheme is based on the hardness of the SIS problem on lattice. The security proof depend on the theory of rejection sampling in[26]. In the random oracle model, our scheme can be proved strongly unforgeable against adaptive chosen messages and ensure unconditional anonymity by rejection sampling. Compared with the lattice-based ring signature scheme proposed in literature [24, 25], our scheme has a very major improvement on enhancing the efficiency, such as the running time, the key size and the signature length. Compared with the traditional lattice-based ring signature scheme, our scheme in the signing process just uses the vector matrix multiplication instead of other algorithms similar to extend base algorithm. Therefore, our scheme has a small amount of operations and the running time is short. In terms of the storage space, our scheme has shorter signature length and saves the signature space consumed in the store.

## 2. DEFINITION

### 2.1 Lattice

**Definition 2.1**(LATTICE[27 DEFINITION1]). Given $n$ linearly independent vectors $\mathbf{b}_1, \mathbf{b}_2, ..., \mathbf{b}_n \in R^m$, the lattice generated by them is defined as

$$L(\mathbf{b}_1, \mathbf{b}_2, ..., \mathbf{b}_n) = \{\sum_{i=1}^{n} x_i \mathbf{b}_i \mid x_i \in Z\}$$

We refer to $\mathbf{b}_1, \mathbf{b}_2, ..., \mathbf{b}_n$ as a basis of the lattice. Equivalently, if we define **B** as the $m \times n$ matrix whose columns are $\mathbf{b}_1, \mathbf{b}_2, ..., \mathbf{b}_n$, the lattice generated by **B** is

$$L(\mathbf{b}_1, \mathbf{b}_2, ..., \mathbf{b}_n) = \{\sum_{i=1}^{n} x_i \mathbf{b}_i \mid x_i \in Z\}$$

We say that the rank of the lattice is $n$ and its dimension is $m$. If $n=m$, the lattice is called a full-rank lattice.

## 2.2 SIS problem

**Definition 2.2** ($l_2 - SIS_{q,n,m,b}$ [26, DEFINITION 3.1]). Given a random matrix $\mathbf{A} \xleftarrow{\$} Z_q^{n \times m}$, find a vector $\mathbf{v} \in Z^m \setminus \{0\}$ such that $\mathbf{Av} = 0$ and $\|\mathbf{v}\| \leq b$.

**Definition 2.3**($SIS_{q,n,m,d}$ distribution [26, DEFINITION 3.2]) choose a matrix $\mathbf{A} \xleftarrow{\$} Z_q^{n \times m}$ and a vector $\mathbf{s} \xleftarrow{\$} \{-d,...0,...,d\}^m$ and output $(\mathbf{A}, \mathbf{As})$.

**Definition 2.4** $SIS_{q,n,m,d}$ search [26, DEFINITION 3.3]. Given a pair $(\mathbf{A},\mathbf{t})$ from the $SIS_{q,n,m,d}$ distribution, find $\mathbf{s} \in \{-d,...0,...,d\}^m$ such that $\mathbf{As} = \mathbf{t}$.

**Definition 2.5** $SIS_{q,n,m,d}$ determination [26, DEFINITION 3.4]. Given a pair $(\mathbf{A},\mathbf{t})$, decide, with non-negligible advantage, whether it came from the $SIS_{q,n,m,d}$ distribution or it was generated uniformly at random from $Z_q^{n \times m} \times Z_q^n$.

## 2.3 The Normal Distribution

Definition 2.6(The continuous Normal distribution [26, DEFINITION 4.1]). The continuous Normal distribution over $R^m$ centered at $\mathbf{v}$ with standard deviation $s$ is defined by the function $r_{\mathbf{v},s}^m(\mathbf{x}) = \left(\dfrac{1}{\sqrt{2ps^2}}\right)^m e^{\frac{-\|\mathbf{x}-\mathbf{v}\|^2}{2s^2}}$. When $\mathbf{v}=0$, we will just write it as $r_s^m(\mathbf{x})$.

Definition 2.7(The discrete Normal distribution[26, DEFINITION 4.2]). The discrete Normal distribution over $Z^m$ centered at some $\mathbf{v} \in Z^m$ with standard deviation $s$ is defined as $D_{\mathbf{v},s}^m(\mathbf{x}) = r_{\mathbf{v},s}^m(\mathbf{x}) / r_s^m(Z^m)$.

The below lemma collects some basic facts about the discrete Normal distribution over $Z^m$.

Lemma 2.1[28-30]:

1. $\Pr[|z| > w(s\sqrt{\log m}); z \xleftarrow{\$} D_s^1] = 2^{-w(\log m)}$

2. For any $\mathbf{z} \in Z^m$, and $s \geq \sqrt{\log 3m}$, $D_s^m(\mathbf{z}) \leq 2^{-m+1}$

3. $\Pr[|\mathbf{z}| > 2s\sqrt{m}; \mathbf{z} \xleftarrow{\$} D_s^m] < 2^{-m}$

# 3. Algorithm

### 3.1 Rejection Sampling

**Lemma 3.1**(rejection sampling [26]). Let $V$ be a subset of $Z^m$ in which all elements have norms less than T, $s$ be some element in R such that $s = w(T\sqrt{\log m})$, and $h: V \to R$ be a probability distribution. Then there exists a constant $M = O(1)$ such that the distribution of the following algorithm $A$:

1. $\mathbf{v} \xleftarrow{\$} h$
2. $\mathbf{z} \xleftarrow{\$} D_{v,s}^m$
3. outputs $(\mathbf{z},\mathbf{v})$ with probability with $\min\left(\dfrac{D_s^m(\mathbf{z})}{MD_{v,s}^m(\mathbf{z})}, 1\right)$

is within statistical distance $\dfrac{2^{-w(\log m)}}{M}$ of the distribution of the following algorithm $F$:

1. $\mathbf{v} \xleftarrow{\$} h$
2. $\mathbf{z} \xleftarrow{\$} D_s^m$
3. outputs $(\mathbf{z}, \mathbf{v})$ with probability with $1/M$

Moreover, the probability that $A$ outputs something is at least $\dfrac{1 - 2^{-w(\log m)}}{M}$.

### 3.2 Trapdoor Function

1. TrapGen($1^n$)[21]. Let $q = poly(n)$ be a prime, $m$ be an arbitrary positive integer such that $m > 5n \log q$. Input a security parameter $n$, and then output the matrix $\mathbf{A} \in Z_q^{n \times m}$ and $\mathbf{B_A} \in Z^{m \times m}$. Here $\mathbf{B_A}$ is a good basis of lattice $\Lambda_q^\perp(\mathbf{A}) = \{\mathbf{v} \in Z^m : \mathbf{Av} = 0 (\bmod q)\}$, and $\|\tilde{\mathbf{B}}_\mathbf{A}\| \leq O(\sqrt{n \log q})$.

2. SamplePr $(\mathbf{A},\mathbf{B_A},s,\mathbf{y})$ [21]. Input $\mathbf{A} \in Z_q^{n \times m}$, $\mathbf{B_A} \in Z^{m \times m}$, any $s \geq \|\mathbf{B}_\mathbf{A}\| \cdot w(\sqrt{\log n})$ and vector $\mathbf{y} \in Z_q^n$. Then obtain a randomly nonzero vector $\mathbf{e} \in \{\mathbf{e} \in Z^m : \|\mathbf{e}\| \leq s\sqrt{m}\}$ such that $\mathbf{Ae} = \mathbf{y}(\bmod q)$ with overwhelming probability.

Literature [21] proposed that the domain of the nonzero vector $\mathbf{e}$ can be defined as $\{\mathbf{e} \in Z^m : \|\mathbf{e}\|_\infty \leq s \cdot w\sqrt{\log n}\}$ in terms of $l_\infty$ norm. Therefore, the SamplePre $(\mathbf{A},\mathbf{B_A},s,\mathbf{y})$ algorithm can be rewritten as follows. Input $\mathbf{A} \in Z_q^{n \times m}$, $\mathbf{B_A} \in Z^{m \times m}$, any $s \geq \|\mathbf{B}_\mathbf{A}\| \cdot w(\sqrt{\log n})$ and vector $\mathbf{y} \in Z_q^n$. Then obtain a randomly nonzero vector $\{\mathbf{e} \in Z^m : \|\mathbf{e}\|_\infty \leq s \cdot w\sqrt{\log n}\}$ by the basis $\mathbf{B_A}$ such that $\mathbf{Ae} = \mathbf{y}(\bmod q)$ with overwhelming probability.

# 4. Ring Signature

## 4.1 Ring Signature

We will describe a ring signature scheme by triple of algorithm (Ring-gen, Ring-sign, Ring-verify).

1. Ring-gen

   The Ring-gen algorithm is a probabilistic polynomial time algorithm that takes as input a security parameter *n* and output a key pair $(pk_i, sk_i)$. Different key pairs may come from different public key systems.

2. Ring-sign

   This algorithm is a probabilistic polynomial time algorithm that takes as input a set of parameters $r$, a signing key $sk_j$, a message $m$, and a set of public keys $L = \{pk_1, ..., pk_l\}$, outputs a ring signature $z$. A certain parameter of $z$ is connected into a ring according the rules. The signer uses the others' public keys to generate a ring with a fracture, and put his private key into the ring to make it become a complete ring.

3. Ring-verify

   The algorithm is a deterministic algorithm that takes as input a set of parameters $r$, a ring signature $z$ on a message $m$, and outputs True or False for accept or reject respectively.

## 4.2 Security of the ring signature scheme

In order to be considered secure, a ring signature scheme must satisfy the properties of both anonymity and unforgeability.

1. Anonymity

   If the probability of an adversary $\Gamma$ wins the anonymity game with the polynomial time is negligible, the ring signature scheme has the anonymity property. The anonymity game is defined as follows.

(1) The challenger $y$ chooses the system parameter *n*, and runs the ring-gen algorithm, then outputs a key pair $\{pk_i, sk_i\}_{(i \in [l])}$. $y$ sends a set of public key $\{pk_i\}_{(i \in [l])}$ to adversary $\Gamma$.

(2) $\Gamma$ asks the ring signature to $y$ for a ring $L = \{\mathbf{A}_i\}_{i \in [l]}$, a message $m$ and an index $j \in [l]$ of the signer. $y$ outputs a ring signature by running the ring-sign algorithm, and sends the result to the adversary $\Gamma$.

(3) The adversary $\Gamma$ adaptability queries the private key of the user whose index is $j$. Then $\mathcal{Y}$ replies with $sk_i$.

(4) $\Gamma$ queries the ring signature with a message $m'$, the index $i_0, i_1 \in [l]$, and a ring $L = \{pk_i\}_{(i \in [l])}$. $\mathcal{Y}$ chooses a random bit $b \in \{0,1\}$, input the private key $sk_{i_b}$ of the user $i_b$, then finishes running ring-sign algorithm and outputs a ring signature $z$.

(5) $\Gamma$ outputs a bit $b'$ as a guess of the user $i_b$. If $b' = b$, $\Gamma$ wins the anonymity game.

2. unforgeability

There is a polynomial-time reduction from the unforgeability of the ring signature scheme to the hard problem. There is a polynomial-time forger **F**, who can query the random oracle $H$ and the Ring-sign algorithm. If **F** can forge a valid ring signature with probability $d$, then there exist a polynomial-time algorithm who can solve the hard problem with some probability. We can prove the unforgeability property by the hardness of the hard problem. The proof consists of the following four steps:

(1) Setup: The challenger $\mathcal{Y}$ designs Hybrid algorithm to replace the ring signing. The advantage that the forger **F** distinguishes the actual Ring-sign algorithm from the one in Hybrid is negligible.

(2) Query Phase: The forger **F**, who can query the Ring-sign algorithm, inputs a message $m$, a ring $L = \{pk_i\}_{(i \in [l])}$ and an index $i_0$ of the signer, then obtains the ring signature. Meanwhile, **F** can also query the random oracle $H$.

(3) Forgery Phase: **F** produces a valid ring signature $(z, c)$ with probability $d$, we will show that **c** comes from query phase with large probability.

(4) Reduction: Discuss the source of **c** in two cases. **c** comes from the query to the ring signing or the random oracle $H$. In each case, we show that there is a polynomial-time algorithm which can solve the hard problem with some probability respect to $d$.

## 5. Lattice-Based Ring Signature Scheme under the Random Oracle Model

In this section we will give our lattice-based ring signature scheme, the scheme is as follows.
1. Setup.

(1) let $q \geq 3$ be a prime number, $n$ a positive integer and large more than 64. $m$ is a positive integer such that $m > 5n \log q$.
$H : \{0,1\}^* \to \{\mathbf{v} : \mathbf{v} \in \{-1,0,1\}^k, \|\mathbf{v}\|_1 \leq k\}$ is collision-resistant hash function, where $k$ and $\mathbf{k}$ are positive integers. A matrix $\mathbf{T}$ is chosen uniformly at random from $Z_q^{n \times k}$.

(2) In a ring $\mathbf{R}$ of $l$ members, for all $i \in [l]$, we run the algorithm TrapGen($1^n$) to obtain $\mathbf{A}_i \in Z_q^{n \times m}$ and $\mathbf{B}_i \in Z_q^{m \times m}$, where $\mathbf{B}_i$ is a good basis of lattice $\Lambda_q^\perp(\mathbf{A}_i) = \{\mathbf{v} \in Z^m : \mathbf{A}_i \mathbf{v} = 0 (\mod q)\}$ and such that $\|\tilde{\mathbf{B}}_i\| \leq L$ with $L = O(\sqrt{n \log q})$.

(3) For all $i \in [l]$, we run the algorithm SamplePre $(\mathbf{A}_i, \mathbf{B}_i, d(w(\sqrt{\log n}))^{-1}, \mathbf{t}_j)$ $k$ times with a positive integer $d \geq \max_{i \in [l]} \{\|\mathbf{B}_i\| \cdot w(\sqrt{\log n})\} (w(\sqrt{\log n}))$. Where vector $\mathbf{t}_j \in Z_q^n (j \in [k])$ is the $j$-th column vector of the matrix $\mathbf{T}$. The algorithm outputs vector $\mathbf{s}_{i,j} \in Z^m (j \in [k])$ such that $\mathbf{A}_i \mathbf{S}_i = \mathbf{T}$ and $\mathbf{S}_i \in \{-d, \cdots, 0, \cdots d\}^{m \times k}$, in which $\mathbf{S}_i = (\mathbf{s}_{i,1}, \cdots, \mathbf{s}_{i,k}) \in \{-d, \cdots, 0, \cdots d\}^{m \times k}$. Let $\mathbf{A}_i \in Z_q^{n \times m}$ be the public key of the $i$-th member, and $\mathbf{S}_i \in \{-d, \cdots, 0, \cdots d\}^{m \times k}$ be the private key associated to $\mathbf{A}_i \in Z_q^{n \times m}$.

2. Sign.

Take as input a message $\mathbf{m}$, a ring $\mathbf{R}$ of $l$ members with public keys $L = \{\mathbf{A}_i\}_{i \in [l]}$, and a private key $\mathbf{S}_j$ of the signer $j$.

(1) For all $i \in [l]$, sampling random vector $\mathbf{y}_i \xleftarrow{\$} D_s^m$, $D_s^m$ is a discrete Normal distribution over $Z^m$ with standard deviation $s$.

(2) set $\mathbf{c} \leftarrow H\left(\sum_{i \in [l]} \mathbf{A}_i \mathbf{y}_i, L, \mathbf{m}\right)$.

(3) For all $i \in [l]$, if $i \neq j$, set $\mathbf{z}_i = \mathbf{y}_i$; if $i = j$, set $\mathbf{z}_j = \mathbf{S}_j \mathbf{c} + \mathbf{y}_j$ with probability $\min\left(\frac{D_s^m(\mathbf{z}_j)}{M D_{\mathbf{S}_j \mathbf{c}, s}^m(\mathbf{z}_j)}, 1\right)$, where $M = O(1)$ is the same as in Lemma 3.1.

(4) Output $(\mathbf{z}_i : i \in [l], \mathbf{c})$ as the ring signature of message $\mathbf{m}$ with ring $\mathbf{R}$ with public keys $L = \{\mathbf{A}_i\}_{i \in [l]}$.

3. Verify.

Take as input a message $m$, a ring $\mathbf{R}$ of $l$ members with public keys $L = \{\mathbf{A}_i\}_{i \in [l]}$, and the ring signature $(\mathbf{z}_i : i \in [l], \mathbf{c})$ of message $m$ with ring $\mathbf{R}$. The verifier accepts the signature if and only if both of the following conditions satisfied:

(1) $\|\mathbf{z}_i\| \leq 2s\sqrt{m}$ for all $i \in [l]$.

(2) $\mathbf{c} = H\left(\sum_{i \in [l]} \mathbf{A}_i \mathbf{z}_i - \mathbf{T}\mathbf{c}, L, m\right)$.

Otherwise, the verifier rejects.

Firstly we will claim that our scheme is correct. Let $(\mathbf{z}_i : i \in [l], \mathbf{c})$ be the signature generated by the user $j$. We will apply the rejection sampling theorem. $\mathbf{S}_j \mathbf{c}$ in the signature algorithm is equivalent to the random vector $\mathbf{v}$ in Lemma 3.1. $\mathbf{z}_j = \mathbf{S}_j \mathbf{c} + \mathbf{y}_j$ is equivalent to $\mathbf{z}$ in Lemma 3.1. According to the conclusions of Lemma 3.1, the distribution of $\mathbf{z}_j = \mathbf{S}_j \mathbf{c} + \mathbf{y}_j$ is within statistical distance $\frac{2^{-w(\log m)}}{M}$ of the Gaussian distribution $D_s^m$. That is to say, the distribution of $\mathbf{z}_j = \mathbf{S}_j \mathbf{c} + \mathbf{y}_j$ is statistical close the Gaussian distribution $D_s^m$. Moreover, in the signature, in addition to the index $j$, $\mathbf{z}_i (i \in [l] \setminus \{j\})$ will all directly come from the Gaussian distribution. Therefore, all $\mathbf{z}_i (i \in [l])$ can be applied to lemma 2.1, $\Pr[|\mathbf{z}| > 2s\sqrt{m}; \mathbf{z} \leftarrow D_s^m] < 2^{-m}$. We have a conclusion that $\|\mathbf{z}_i\| \leq 2s\sqrt{m} (i \in [l])$ holds with an overwhelming probability. The first part of the verification is correct for a valid signature.

With respect to the second test, we have:

$$\begin{aligned}
\sum_{i \in [l]} \mathbf{A}_i \mathbf{z}_i - \mathbf{T}\mathbf{c} &= \mathbf{A}_j \mathbf{z}_j - \mathbf{T}\mathbf{c} + \sum_{i \in [l] \setminus \{j\}} \mathbf{A}_i \mathbf{z}_i \\
&= \mathbf{A}_j \mathbf{S}_j \mathbf{c} + \mathbf{A}_j \mathbf{y}_j - \mathbf{T}\mathbf{c} + \sum_{i \in [l] \setminus \{j\}} \mathbf{A}_i \mathbf{y}_i \\
&= \sum_{i \in [l]} \mathbf{A}_i \mathbf{y}_i
\end{aligned}$$

Therefore, from the above argument we can conclude that our ring signature scheme is correct.

# 6. Anonymity

**Theorem 6.1**(Anonymity): The probability of an adversary won the anonymity game with the polynomial time is negligible. Therefore, the ring signature scheme is of anonymity.

Proof: Given a challenger $y$, and an adversary $\Gamma$, consider the following game:

1. The challenger $y$ obtain $\mathbf{A}_i \in Z_q^{n \times m}$, and $\mathbf{B}_i \in Z_q^{m \times m}$ ($i \in [l]$) that is a good basis of lattice $\Lambda_q^{\perp}(\mathbf{A}_i)$, by $l$ times of running $\text{TrapGen}(1^n)$, and matrix $\mathbf{T}$ that is chosen uniformly at random from $Z_q^{n \times k}$. Then outputs matrix $\mathbf{S}_i \in \{-d, \cdots, 0, \cdots d\}^{m \times k}$, which are generated by running $k$ times of SamplePre $(\mathbf{A}_i, \mathbf{B}_i, d(w(\sqrt{\log n}))^{-1}, \mathbf{t}_j)$, so it satisfies $\mathbf{A}_i \mathbf{S}_i = \mathbf{T}$. The challenger $y$ gives public key $\mathbf{A}_i$ for all $i \in [l]$ and the parameter $\mathbf{T}$ is given to $\Gamma$, keep private key $\mathbf{S}_i$ for all $i \in [l]$ secretly.

2. $y$ answers the signing query of adversary $\Gamma$, with a ring $L = \{\mathbf{A}_i\}_{i \in [l]}$, and a message $m$. $y$ sends the results of the ring-sign to the adversary $\Gamma$.

3. The adversary $\Gamma$ adaptability queries the private key of the index $i \in [l]$, $y$ reply with $\mathbf{S}_i$.

4. The adversary $\Gamma$ queries the ring signature with a message $m'$, the index $i_0, i_1 \in [l]$, and the public key $L = \{\mathbf{A}_i\}_{i \in [l]}$ of a ring. $y$ chooses a random bit $b \in \{0,1\}$, inputs the private key $\mathbf{S}_{i_b}$ of the user $i_b$, then finishes running ring-sign and outputs a ring signature $(z_i : i \in [l], c)$.

5. $\Gamma$ outputs a bit $b'$ as a guess of the random bit $b \in \{0,1\}$.

If we running the ring-sign algorithm by the private key $\mathbf{S}_{i_b}$ of the index $i_b$, we will get $\mathbf{z}_{i_b} = \mathbf{S}_{i_b} \mathbf{c} + \mathbf{y}_{i_b}$ with probability $\min\left(\dfrac{D_s^m(\mathbf{z}_{i_b})}{M D_{\mathbf{S}_{i_b} \mathbf{c}, s}^m(\mathbf{z}_{i_b})}, 1\right)$, and output the sign $(\mathbf{z}_i : i \in [l], \mathbf{c})$. And if we use another method to obtain $\mathbf{z}_{i_b}$ in the ring-sign algorithm, such as that we obtain $\mathbf{z}_{i_b}$ that are chosen uniformly at random over $D_s^m$ with probability $1/M$ ($M = O(1)$), and output a ring signature $(\mathbf{z}_i : i \in [l], \mathbf{c})$. Using the rejection sampling lemma 3.1, the statistical distance of the distribution of the signature from those two algorithms is within $\frac{2^{-w(\log m)}}{M}$.

Let $X_{i_b}$ be a distribution that represents the distribution of the signature received by the private key $\mathbf{S}_{i_b}$, while $\mathbf{z}_{i_b}$ is obtained uniformly at random over $D_\mathbf{s}^m$ with probability $1/M$. And the remaining $\mathbf{z}_i (i \in [l] \setminus i_b)$ come from $D_\mathbf{s}^m$, we use $Y$ to denote the distribution of the signature $(\mathbf{z}_i : i \in [l], \mathbf{c})$. Using the rejection sampling lemma 3.1 for $i_0$ $i_1$, and use above analysis

$$\Delta(X_{i_0}, X_{i_1}) \leq \Delta(X_{i_0}, Y) + \Delta(X_{i_1}, Y) \leq 2 \cdot \frac{2^{-w(\log m)}}{M}$$

Therefore, the signature $X_{i_0}$ is indistinguishable with the signature $X_{i_1}$ with overwhelming probability. That is to say, the advantage of any adversary making a correct guess of $b$ is negligible. This implies the ring signature scheme ensures unconditional anonymity.

## 7. Unforgeability

Before proving the unforgeability, we need to prove the following theorem, there exists a collision of the vector $\mathbf{s}$ with certain probability.

**Theorem 6.2** Given any $\mathbf{A} \in Z_q^{n \times m}$ where $n \gg 64$ and $m > 5n \log q$, randomly chosen $\mathbf{s} \xleftarrow{\$} \{-d, \cdots, 0, \cdots, d\}^m$, then with probability $1 - 2^{-100}$, there exists another $\mathbf{s}' \xleftarrow{\$} \{-d, \cdots, 0, \cdots, d\}^m$ such that $\mathbf{As} = \mathbf{As}'$.

Proof: Similar to the proof of Theorem 5.2 in the literature [26]. Notice that $\mathbf{A}$ can be thought of as a linear transformation whose range has size $q^n$. This means that there are at most $q^n$ elements $\mathbf{s} \xleftarrow{\$} \{-d, \cdots, 0, \cdots, d\}^m$ that do not collide with any other elements in $\{-d, ..., 0, ..., d\}^m$. Since the set $\{-d, ..., 0, ..., d\}^m$ consists of $(2d+1)^m$ elements, the probability of randomly selecting a non-colliding element is at most

$$\frac{q^n}{(2d+1)^m} \leq \frac{q^n}{(2d+1)^{5n \log q}} \leq \frac{q^n}{(2d+1)^{64 + n \log q / \log(2d+1)}} = \frac{1}{(2d+1)^{64}} < 2^{-100}$$

In the process of proving unforgeability, we will use the following algorithms Hybrid1 and Hybrid2. When the adversary asks to the signature, the challenger can use the algorithm Hybrid1 or Hybrid2 to replace the real signature algorithm, and reply to adversary's signature queries. The following theorem can prove that the statistical distance between the algorithm Hybrib1 or Hybrid2 and the real signature algorithm is negligible.

Hybrid1 $\left(m, L = \{\mathbf{A}_i\}_{i \in [l]}, j\right)$

1. For all $i \in [l]$, sample random vector $\mathbf{y}_i \xleftarrow{\$} D_s^m$
2. $c \xleftarrow{\$} \{\mathbf{v}: \mathbf{v} \in \{-1,0,1\}^k, \|\mathbf{v}\|_1 \leq k\}$
3. For all $i \in [l]$, if $i \neq j$, set $\mathbf{z}_i = \mathbf{y}_i$; if $i = j$, set $\mathbf{z}_j = \mathbf{S}_j \mathbf{c} + \mathbf{y}_j$ with

    probability $\min\left(\dfrac{D_s^m(\mathbf{z}_j)}{M D_{\mathbf{S}_j \mathbf{c}, s}^m(\mathbf{z}_j)}, 1\right)$

4. Output $(\mathbf{z}_i : i \in [l], \mathbf{c})$
5. Program $H\left(\sum_{i \in [l]} \mathbf{A}_i \mathbf{z}_i - \mathbf{Tc}, L, m\right) = \mathbf{c}$

Hybrid2 $\left(m, L = \{\mathbf{A}_i\}_{i \in [l]}, j\right)$

1. $c \xleftarrow{\$} \{\mathbf{v}: \mathbf{v} \in \{-1,0,1\}^k, \|\mathbf{v}\|_1 \leq k\}$
2. $\mathbf{z}_i \xleftarrow{\$} D_s^m$. For all $i \in [l]$, $\mathbf{z}_i \xleftarrow{\$} D_s^m$
3. with probability $\dfrac{1}{M}$, Output $(\mathbf{z}_i : i \in [l], \mathbf{c})$
4. Program $H\left(\sum_{i \in [l]} \mathbf{A}_i \mathbf{z}_i - \mathbf{Tc}, L, m\right) = \mathbf{c}$

**Theorem 6.3**. Suppose there is a polynomial-time forger **F**, who makes at most $s$ queries to the signing oracle and $h$ queries to the random oracle $H$, succeeds in forging the ring signature scheme with non-negligible probability $d$, then there is a polynomial-time algorithm who can solve the $l_2 - SIS_{q,n,ml,b}$ problem for $b = (4s + 2dk)\sqrt{ml}$ with non-negligible probability $\approx \dfrac{d^2}{2(h+s)}$.

Proof: The theorem is proved by theorem 6.4 and theorem 6.5. In theorem 6.4, we will show that our ring signing algorithm can be replaced by Hybrid2, and the statistical distance between the two outputs will be at most $e = s(h+s) \cdot 2^{-nl+1} + s \cdot \dfrac{2^{-100}}{M}$. In theorem 6.5, we will show that if a forger can produce a forgery with non-negligible probability $d$ when the ring signing algorithm is replaced by one in Hybrid2, we can utilize him to recover a vector $\mathbf{v}$ such that $\mathbf{Av}=0$ and $\|\mathbf{v}\| \leq (4\mathbf{s}+2d\mathbf{k})\sqrt{ml}$, with non-negligible probability at least

$$\left(\frac{1}{2} - 2^{-100}\right)\left(d - \frac{1}{|D_H|}\right)\left(\frac{d - 1/|D_H|}{t} - \frac{1}{|D_H|}\right)$$

If we assume that the output of the hash function $H$ is 100 bits in the ring signing scheme, the probability is $(\frac{1}{2} - 2^{-100})(d - 2^{-100})(\frac{d - 2^{-100}}{h+s} - 2^{-100}) \approx \dfrac{d^2}{2(h+s)}$.

**Theorem 6.4** Let $D$ be a distinguisher who can query the random oracle $H$ with $h$ times and Hybrid2 with $s$ times. For all but a $e^{-\Omega(n)}$ fraction of all possible matrices $\mathbf{A}_i\ i \in [l]$, the advantage that $D$ distinguishes the actual ring signing algorithm from the one in Hybrid2 is at most $s(h+s) \cdot 2^{-nl+1} + s \cdot \dfrac{2^{-w(\log m)}}{M}$.

Proof: We first show that the Distinguisher $D$ has advantage of at most $s(h+s) \cdot 2^{-nl+1}$ in distinguishing between the real ring signing algorithm and Hybrid1. The only difference between the two algorithm is that in Hybrid1, instead of the output from the hash function $H$, vector $\mathbf{c}$ is chosen at uniformly random from $\{\mathbf{v}: \mathbf{v} \in \{-1,0,1\}^k, \|\mathbf{v}\|_1 \leq k\}$. And then the procedure sets the answer with $H\left(\sum_{i \in [l]} \mathbf{A}_i \mathbf{z}_i - \mathbf{Tc}, L, \mathbf{m}\right) = H\left(\sum_{i \in [l]} \mathbf{A}_i \mathbf{y}_i, L, \mathbf{m}\right)$ without checking whether the value for $\left(\sum_{i \in [l]} \mathbf{A}_i \mathbf{y}_i, L, \mathbf{m}\right)$ was already set. Since $D$ calls the random oracle $H$ $h$ times and the ring signing algorithm $s$ times, at most $s+h$ values of $\left(\sum_{i \in [l]} \mathbf{A}_i \mathbf{y}_i, L, \mathbf{m}\right)$ have ever been set. We now show that each time the Hybrid1 procedure is called, the probability of generating $l$ vectors $y_i\ i \in [l]$ such that $\sum_{i \in [l]} \mathbf{A}_i \mathbf{y}_i$ is equal to one of the previous queried values is at most of $2^{-nl+1}$. With probability $1 - e^{-\Omega(n)}$, write the matrix $\mathbf{A}_i\ i \in [l]$ as "Hermite Normal Form", such as $\mathbf{A}_i = [\overline{\mathbf{A}}_i \| \mathbf{I}]$, and write the corresponding vector $\mathbf{y}_i$ as $\mathbf{y}_i = [\mathbf{y}_{i,1} \| \mathbf{y}_{i,0}]$. For any $\mathbf{t} \in Z_q^n$,

$$\Pr\left[\sum_{i\in[l]} \mathbf{A}_i \mathbf{y}_i = \mathbf{t}; \mathbf{y}_i \xleftarrow{\$} D_s^m\right]$$

$$= \Pr\left[\sum_{i\in[l]} \mathbf{y}_{i,1} = \left(\mathbf{t} - \sum_{i\in[l]} \overline{\mathbf{A}}_i \mathbf{y}_{i,0}\right); \mathbf{y}_i \xleftarrow{\$} D_s^m\right]$$

$$= \Pr\left[\mathbf{y}_1 = [\mathbf{y}_{1,1} \| \cdots \| \mathbf{y}_{l,1}] = \left(\mathbf{t} - \sum_{i\in[l]} \overline{\mathbf{A}}_i \mathbf{y}_{i,0}\right); \mathbf{y}_i \xleftarrow{\$} D_s^m\right]$$

$$= \max_{\mathbf{t}' \in Z_q^{nl}} \Pr\left[\mathbf{y}_1 = \mathbf{t}'; \mathbf{y}_1 \xleftarrow{\$} D_s^{nl}\right]$$

$$\leq 2^{-nl+1}$$

Where the last inequality follows Lemma 2.1. Thus if Hybrib1 is accessed $s$ times, and the probability of getting a collision each time is at most $(h+s) \cdot 2^{-nl+1}$, the probability that a collision occurs after $s$ queries is at most $s(h+s) \cdot 2^{-nl+1}$.

Using rejection sampling lemma 3.1, Hybrid1 plays the role of the algorithm $A$ in lemma 3.1, and Hybrid2 corresponds to $F$, the statistical distance between the output of Hybrid1 and the output of Hybrid2 is within $\dfrac{2^{-w(\log m)}}{M}$. Since the ring signing is called at most $s$ times, the statistical distance is at most $s \cdot \dfrac{2^{-w(\log m)}}{M}$.

In conclusion, the statistical distance between the real ring signing algorithm and Hybrid2 is within

$$s(h+s) \cdot 2^{-nl+1} + s \cdot \dfrac{2^{-w(\log m)}}{M}$$

That is the advantage of distinguishing between the real ring signing algorithm and Hybrid2.

**Theorem 6.5** Suppose there exists a polynomial-time forger $\mathbf{F}$, who makes at most $s$ queries to the signing oracle and $h$ queries to the random oracle $H$, succeeds in forging the ring signature with non-negligible probability $d$, then there exists an algorithm of the same time-complexity as $\mathbf{F}$, given $\mathbf{A} = [\mathbf{A}_1 \| \cdots \| \mathbf{A}_l] \xleftarrow{\$} Z_q^{n \times ml}$, the algorithm can find a non-zero $\mathbf{v} \in Z^{ml}$ such that $\mathbf{A}\mathbf{v} = 0$ and $\|\mathbf{v}\| \leq (4s + 2d\mathbf{k})\sqrt{ml}$, with non-negligible probability at least

$$(\frac{1}{2} - 2^{-100})(d - 2^{-100})(\frac{d - 2^{-100}}{h+s} - 2^{-100})$$

Proof: Let $D_H = \{\mathbf{c} : \mathbf{c} \in \{-1,0,1\}^k, \|\mathbf{c}\|_1 \leq k\}$ denotes the range of the random oracle $H$. Given a randomly matrix $\mathbf{T}$ over $Z_q^{n \times k}$, let $t=h+s$ be the number of times the random oracle $H$ is called or programmed during $\mathbf{F}$'s attack. We note $r_1,...,r_t \xleftarrow{\$} D_h$ as the responses of the random oracle $H$.

We now consider a subroutine $A$, which takes as input $(\{\mathbf{A}_i\}_{i \in [l]}, \mathbf{T}, i_0, r_1,...,r_t)$, where $i_0 \in [l]$ is an index of the signer. We will initialize the forger $\mathbf{F}$ by giving the parameter $(\{\mathbf{A}_i\}_{i \in [l]}, \mathbf{T}, i_0)$. When $\mathbf{F}$ wants to get a signature of the message $\mathbf{m}$, $A$ runs the ring signing algorithm in Hybrid2 $(\mathbf{m}, L = \{\mathbf{A}_i\}_{i \in [l]}, i_0)$ to produce the signature. During signing, the random oracle $H$ will have to be programmed, and the response of $H$ will be first $r_i (i \in [t])$ in the list $(r_1,...,r_t)$ that hasn't been used yet. Therefore, $A$ will have to keep a table of all the queries to $H$. In case the same query is made twice, it will have to reply with the previously answer $r_i$. The forger $\mathbf{F}$ can also make queries to the random oracle $H$, in which case the reply will similarly be the first unused $r_i$ in the list $(r_1,...,r_t)$. Once $\mathbf{F}$ finishes running and outputs a forgery with non-negligible probability $d$, our subroutine $A$ simply outputs $\mathbf{F}$'s output.

With probability $d$, $\mathbf{F}$ will output a valid ring signature $(\mathbf{z}_i : i \in [l], \mathbf{c})$ for $(\mathbf{m}, L = \{\mathbf{A}_i\}_{i \in [l]}, i_0)$, then we have $\|\mathbf{z}_i\| \leq 2s\sqrt{m}$ and $\mathbf{c} = H\left(\sum_{i \in [l]} \mathbf{A}_i \mathbf{z}_i - \mathbf{Tc}, L, \mathbf{m}\right)$ for all $i \in [l]$. Notice that if the random oracle $H$ was not queried or programmed on some input $\mathbf{w} = \sum_{i \in [l]} \mathbf{A}_i \mathbf{z}_i - \mathbf{Tc}$ during forging, then $\mathbf{F}$ only has a $1/|D_H|$ chance to produce vector $\mathbf{c}$ such that $\mathbf{c} = H(\mathbf{w}, L, \mathbf{m})$. To ensure the validity of the ring signature, with probability $1 - 1/|D_H|$, the vector $\mathbf{c}$ must be one of the list $(r_1,...,r_t)$ that is the responses of the random oracle $H$. In conclusion, the probability that $\mathbf{F}$ succeeds in a forgery and $\mathbf{c}$ is one of the list $(r_1,...,r_t)$ is at least

$$d(1 - 1/|D_H|) = d - d/|D_H| \geq d - 1/|D_H|$$

Let $j$ be some index such that $\mathbf{c} = r_j$. There are two possibilities: $r_j$ was a response to a random oracle H query made by **F**, or was programmed during signing.

Case 1: $r_j$ was programmed during signing

The signer programmed the random oracle $H\left(\sum_{i\in[l]} \mathbf{A}_i \mathbf{z}'_i - \mathbf{T}\mathbf{c}, L, \mathbf{m}'\right) = \mathbf{c}$ for signing a message $\mathbf{m}'$. If the forger outputs a valid forgery $(\mathbf{z}_i : i \in [l], \mathbf{c})$ for the message $\mathbf{m}$, we have

$$H\left(\sum_{i\in[l]} \mathbf{A}_i \mathbf{z}'_i - \mathbf{T}\mathbf{c}, L, \mathbf{m}'\right) = H\left(\sum_{i\in[l]} \mathbf{A}_i \mathbf{z}_i - \mathbf{T}\mathbf{c}, L, \mathbf{m}\right)$$

If $\sum_{i\in[l]} \mathbf{A}_i \mathbf{z}'_i - \mathbf{T}\mathbf{c} \neq \sum_{i\in[l]} \mathbf{A}_i \mathbf{z}_i - \mathbf{T}\mathbf{c}$ or $\mathbf{m}' \neq \mathbf{m}$, it means that **F** has found a pre-image of $r_j$. Therefore, we have $\sum_{i\in[l]} \mathbf{A}_i \mathbf{z}'_i - \mathbf{T}\mathbf{c} = \sum_{i\in[l]} \mathbf{A}_i \mathbf{z}_i - \mathbf{T}\mathbf{c}$ and $\mathbf{m}' = \mathbf{m}$. $\sum_{i\in[l]} \mathbf{A}_i \mathbf{z}'_i - \mathbf{T}\mathbf{c} = \sum_{i\in[l]} \mathbf{A}_i \mathbf{z}_i - \mathbf{T}\mathbf{c}$, so $\sum_{i\in[l]} \mathbf{A}_i (\mathbf{z}'_i - \mathbf{z}_i) = 0$. By concatenating each $\mathbf{A}_i, \mathbf{z}_i (i \in [l])$, we have $\mathbf{A}\mathbf{v} = 0$, where $\mathbf{A} = [\mathbf{A}_1 \| \cdots \| \mathbf{A}_l] \in Z_q^{n \times ml}$, $\mathbf{v} = [(\mathbf{z}'_1 - \mathbf{z}_1) \| \cdots \| (\mathbf{z}'_l - \mathbf{z}_l)]$. According to the conditions in the verification algorithm, we have $\|\mathbf{z}'_i\|, \|\mathbf{z}_i\| \leq 2s\sqrt{m}$, hence $\|\mathbf{v}\| \leq 4s\sqrt{ml}$.

Case 2: $r_j$ was a response to a random oracle H

We record the signature $(\mathbf{z}_i : i \in [l], r_j)$ of **F** on the message $\mathbf{m}$, and generate fresh random elements $(r'_j, ..., r'_t) \leftarrow D_H$. We then run the subroutine $A$ again with input $(\{\mathbf{A}_i\}_{i\in[l]}, \mathbf{T}, i_0, r_1, ..., r_{j-1}, r'_j, ..., r'_t)$. By the General Forking Lemma, we obtain that the probability that $r'_j \neq r_j$ and the forger uses the random oracle response $r'_j$ in its forgery is at least

$$\left(d - \frac{1}{|D_H|}\right)\left(\frac{d - 1/|D_H|}{t} - \frac{1}{|D_H|}\right).$$

Thus with the above probability, the forger **F** outputs a valid ring signature $(\mathbf{z}'_i : i \in [l], r'_j)$ of the message $\mathbf{m}$, and

$$\sum_{i\in[l]} \mathbf{A}_i \mathbf{z}'_i - \mathbf{T}\mathbf{c}' = \sum_{i\in[l]} \mathbf{A}_i \mathbf{z}_i - \mathbf{T}\mathbf{c}$$

where $\mathbf{c} = r_j$ and $\mathbf{c}' = r'_j$. By rearranging, we obtain

$$\sum_{i\in[l]} \mathbf{A}_i[\mathbf{z}'_i - \mathbf{z}_i + \mathbf{S}_i\mathbf{c}' - \mathbf{S}_i\mathbf{c}] = 0.$$

Let $\mathbf{A} = [\mathbf{A}_1 \| \cdots \| \mathbf{A}_l] \in Z_q^{n \times ml}$, $\mathbf{v} = [(\mathbf{z}'_1 - \mathbf{z}_1 + \mathbf{S}_1\mathbf{c}' - \mathbf{S}_1\mathbf{c}) \| \cdots \| (\mathbf{z}'_l - \mathbf{z}_l + \mathbf{S}_l\mathbf{c}' - \mathbf{S}_l\mathbf{c})]$, we have $\mathbf{Av} = 0$. Since $\|\mathbf{z}'_i\|, \|\mathbf{z}_i\| \leq 2s\sqrt{m}$, and $\|\mathbf{S}_i\mathbf{c}'\|, \|\mathbf{S}_i\mathbf{c}\| \leq dk\sqrt{m}$, we know that $\|\mathbf{v}\| \leq (4s + 2dk)\sqrt{ml}$.

Now all we need is to prove $\mathbf{v} \neq 0$. For any index $\mathbf{z}'_{i_0} - \mathbf{z}_{i_0} + \mathbf{S}_{i_0}\mathbf{c}' - \mathbf{S}_{i_0}\mathbf{c}(i_0 \in [l])$, let $k$ be a position in which $c[k] \neq c'[k]$. By theorem 6.2, we know that there is at least a $1-2^{-100}$ chance that there exists another secret key $\mathbf{S}'_{i_0}$ such that all the columns of $\mathbf{S}'_{i_0}$ are the same as $\mathbf{S}_{i_0}$ except for column $k$, and $\mathbf{AS}'_{i_0} = \mathbf{AS}_{i_0}$. If $\mathbf{z}'_{i_0} - \mathbf{z}_{i_0} + \mathbf{S}_{i_0}\mathbf{c}' - \mathbf{S}_{i_0}\mathbf{c} = 0$, then $\mathbf{z}'_{i_0} - \mathbf{z}_{i_0} + \mathbf{S}'_{i_0}\mathbf{c}' - \mathbf{S}'_{i_0}\mathbf{c} \neq 0$. But to forger $\mathbf{F}$, he dose not know whether we use the secret key $\mathbf{S}'_{i_0}$ or $\mathbf{S}_{i_0}$, the probability that we obtain $\mathbf{z}'_{i_0} - \mathbf{z}_{i_0} + \mathbf{S}_{i_0}\mathbf{c}' - \mathbf{S}_{i_0}\mathbf{c} \neq 0$ is at lest 1/2, and so we will find a non-zero vector $\mathbf{v}$ with probability at least

$$\left(\frac{1}{2} - 2^{-100}\right)\left(d - \frac{1}{|D_H|}\right)\left(\frac{d - 1/|D_H|}{t} - \frac{1}{|D_H|}\right)$$

## 8. Efficiency Analysis

The proposed scheme required just a few matrix-vector multiplications in signing algorithm, which is different with the others who use the ExtBasis algorithm. Therefore, our new scheme is more simple, more efficient, and with shorter signature length. In terms of the storage space, the signature is composed of $l+1$ vectors. Among these, $l$ vectors' length is $m$, so the length of the signtature is $lm+k$. Compared to the other schemes, we have made great progress.

Here in table 8.1 we have compared the efficiency of our scheme with other schemes based on lattices [24, 25] in time and space. Where $T_{Sam}$ and $T_{Ext}$ represent the time cost by algorithm SamplePre and ExtBasis every step, respectively. $T_{mul}$ represents the time cost by scalar multiplication computation $n$ times.

Table 8.1 efficiency comparing

| scheme | Sign time | Verify time | Signature length |
|---|---|---|---|
| 2010W [24] | $m(l+d) T_{Ext} + m(l+d+1) T_{Sam}$ | $m(l+d+1) T_{mul}$ | $(l+d+1)m$ |
| 2012T [25] | $m T_{Sam} + m(l+1) T_{mul}$ | $m(l+2) T_{mul}$ | $(l+2)m$ |
| Our scheme | $m(l+1) T_{mul}$ | $m(l+1) T_{mul}$ | $lm+k$ |

# 9. Conclusion

We present a high efficient lattice-based ring signature scheme, and prove its security of unforgeability in the random oracle model. Our scheme has made a great progress in time efficiency. The length of the signature is also relative improved comparing with others. However, both the length of the public key and private key are longer than the scheme which was presented by Carlos Aguilar-Melchor in 2013. So the storage cost of the public key and private key is relatively high. Therefore, our work in the future is to transform our scheme to polynomial rings in order to improve the length of the public key and private key.

# Acknowledgements

This work is supported by the National Natural Science Foundation of China under grants 61173192, Research Foundation of Education Department of Shaanxi Province of China under grants 12JK0740. Thanks also go to the anonymous reviewers for their useful comments.